\documentclass[namedreferences]{SolarPhysics}
\usepackage[optionalrh]{spr-sola-addons} 
\usepackage{graphicx}        
\usepackage{color}           
\usepackage{url}             

\newcommand{\jasr}{    {\it Adv. Space Res.}}

\newcommand{\aap}{    {\it Astron. Astrophys.}}

\newcommand{\apj}{    {\it Astrophys. J.}}
\newcommand{\apjl}{   {\it Astrophys. J. Lett.}}
\newcommand{\apjs}{   {\it Astrophys. J. Suppl.}}

\newcommand{\grl}{    {\it Geophys. Res. Lett.}}

\newcommand{\jgr}{    {\it J. Geophys. Res.}}

\newcommand{\pasj}{   {\it Publ. Astron. Soc. Japan}}

\newcommand{\solphys}{{\it Solar Phys.}}

\newcommand{\ssr}{    {\it Space Sci. Rev.}}

\begin{document}

\begin{article}

\begin{opening}

\title{Acceleration of Relativistic Protons during the 20 January 2005 Flare and CME}

\author{S.~\surname{Masson}$^{1}$\sep
        K.-L.~\surname{Klein}$^{1}$\sep
        R.~\surname{B\"utikofer}$^{2}$\sep
        E.~\surname{Fl\"uckiger}$^{2}$\sep
        V.~\surname{Kurt}$^{3}$\sep
	B.~\surname{Yushkov}$^{3}$\sep
          S.~\surname{Krucker}$^{4}$
       }
\runningauthor{Masson et al.}
\runningtitle{Acceleration of Relativistic Protons during the 20 January 2005 Flare and CME}

   \institute{$^{1}$ Observatoire de Paris, LESIA - CNRS UMR 8109, 
   	Universit\'es P\&M Curie et Denis Diderot, Paris, 
		  F-92195 Meudon C\'edex, 
                     France 
                     email: \url{sophie.masson@obspm.fr} 
                     email: \url{ludwig.klein@obspm.fr} \\ 
              $^{2}$ University of Bern,  
              	Space Research \& Planetary Sciences,		
                     CH-3012 Bern, 
                     Switzerland
                     email: \url{rolf.buetikofer@space.unibe.ch} \\
                     email: \url{flueckiger@space.unibe.ch} \\
             $^{3}$ Skobeltsyn Institute of Nuclear Physics, 
		    Lomonosov Moscow State University
		    Moscow 119991, Russia
                     email: \url{vgk@srd.sinp.msu.ru} \\
              $^{3}$ Space Sciences Laboratory,
		University of California,
		Berkeley, CA 94720-7450, USA
                   email : \url{krucker@ssl.berkeley.edu}\\
             }

%

%

\begin{abstract}
 The origin of relativistic solar protons during large flare/CME events has not been uniquely identified so far. We perform a detailed comparative analysis of the time profiles of relativistic protons detected by the worldwide network of neutron monitors at Earth with electromagnetic signatures of particle acceleration in the solar corona during the large particle event of 20 January 2005. The intensity-time profile of the relativistic protons derived from the neutron monitor data indicates two successive peaks. We show that microwave, hard X-ray and $\gamma$-ray emissions display several episodes of particle acceleration within the impulsive flare phase. The first relativistic protons detected at Earth are accelerated together with relativistic electrons and with protons that produce pion decay $\gamma$-rays during the second episode. The second peak in the relativistic proton profile at Earth is accompanied by new signatures of particle acceleration in the corona within $\approx 1~R_\odot$ above the photosphere, revealed by hard X-ray and microwave emissions of low intensity, and by the renewed radio emission of electron beams and of a coronal shock wave. We discuss the observations in terms of different scenarios of particle acceleration in the corona.
\end{abstract}

\end{opening}

\section{Introduction}
\label{s-intro} 

Energy conversion in the solar corona is often accompanied by particle acceleration, which on occasion extends to relativistic energies. When relativistic ions or nucleons impinge on the Earth's atmosphere, they produce air showers of, among others, neutrons, protons and muons. The nucleons can be detected by ground based neutron monitors, provided the primary particle at the top of the atmosphere has an energy above some threshold, typically 450~MeV. Transient flux enhancements of relativistic solar particles are called ground level enhancements (GLEs). They are the relativistic extension of solar energetic particle (SEP) events. An overview of neutron monitor studies of GLEs is given by \inlinecite{Lop-06}.

It is still an open question how the Sun accelerates particles, and more specifically, how it can accelerate particles to the relativistic energies which are observed during the GLE. With flares and coronal shock waves, which both accompany large SEP events \cite{Gop:al-04}, solar activity provides candidate environments for particle acceleration. But observations have so far not been able to show unambiguously which of them is the key element for particle acceleration to relativistic energies. 

The links of particles detected near 1~AU with their solar origin are blurred by their propagation in interplanetary space, through scattering by the turbulent magnetic field and reflection at large-scale magnetic structures \cite{Mey:al-56,Dro-00,Bie:al-02,Sai:al-08}. Thus, comparing signatures of accelerated solar particles at the Sun with the measurements of the relativistic particles at the Earth is often difficult, unless particularly favourable conditions of interplanetary propagation are met. This is why detailed timing analyses of individual events observed under such favourable conditions are valuable means to search for common signatures of particle acceleration in GLE time profiles and in electromagnetic signatures at the Sun, especially at $\gamma$-ray, hard X-ray and radio wavelengths. Such studies have been carried out in the past, and suggested that the neutron monitor time profiles give information on coronal acceleration processes \cite{Aki:al-96,Deb:al-97,Kle:al-99a,Kle:al-01,Mir:al-05}. However, the timing uncertainty due to interplanetary propagation makes a clear association of a GLE feature with a specific phase of the flare/CME event in the corona difficult.

The GLE on 20 January 2005 displays a conspicuous and rapid increase of the relativistic particle flux above the cosmic ray background detected by neutron monitors. The prompt increase and the high anisotropy suggest that the time profiles suffered little distortion by interplanetary scattering. In this paper we report on a detailed timing analysis of relativistic protons at 1~AU and interacting particles. Section~\ref{ss-NMs} describes how the time profiles and distribution functions of protons detected at the Earth are derived from a set of neutron monitor measurements. The timing of accelerated particles in the chromosphere and the low corona is inferred from their microwave, hard X-ray and $\gamma$-ray emissions (Section~\ref{ss-chrom}). Metric-to-kilometric radio emission from electron beams is used to trace the injection and propagation of accelerated particles in the corona and interplanetary space (Section~\ref{ss-wind}). The combination of these analyses allows us to infer the interplanetary path length travelled by the protons and their initial solar release time (Section~\ref{ss-obs}), and the relationship of escaping protons with coronal acceleration processes throughout the event (Section~\ref{ss-time}). Finally, we present in Section~\ref{s-disc} a consistent scenario of coronal acceleration in different episodes with different radiative signatures of energetic particles and different conditions of access to interplanetary space. We discuss our findings with respect to other work on this event, and to particle acceleration scenarios of GLEs.

\section{Observations}
\label{s-obs} 

The relativistic proton event that started on 20 January 2005 at 06:50~UT was part of a large SEP event detected by a fleet of spacecraft and by the worldwide network of neutron monitors. Exceptionally energetic particles were detected by muon telescopes on the Earth \cite{Rya:Mil-05,DAn:Poi-05}, which are sensitive to energies above several GeV. The event accompanied intense activity at the Sun comprising a strong flare in soft X-rays (GOES class X7.1, 06:37-07:26~UT, peak at 07:01~UT) and H$\alpha$ (2B, N12$^\circ$ W58$^\circ$, active region NOAA 10720), as reported in {\it Solar Geophysical Data} Comprehensive Reports for January 2005\footnote{NOAA Solar-Terrestrial Physics Division, http://sgd.ngdc.noaa.gov/sgd/jsp/solarfront.jsp}, along with intense hard X-ray and $\gamma$-ray emissions from high-energy particles \cite{Kuz:al-08,Sld:al-08,Kru:al-08}. SOHO observed a broad and fast CME and large-scale disturbances in EUV. Detailed presentations of this activity can be found, {\it e.g.}, in \inlinecite{Sim-06} and \inlinecite{Grc:al-08}. The 20 January 2005 event was one of a series of large flares and CMEs involving the same active region between 13 January 2005 and its passage across the west limb. These events left the interplanetary medium in a highly perturbed state \cite{Pln:al-07,McC:al-08}. Neutron monitors detected a Forbush decrease starting on 17~January \cite{Flu:al-05}.

%
%
\begin{figure}
\centerline{
  \includegraphics[width=0.41\textwidth,height=0.6\textheight,angle=90]{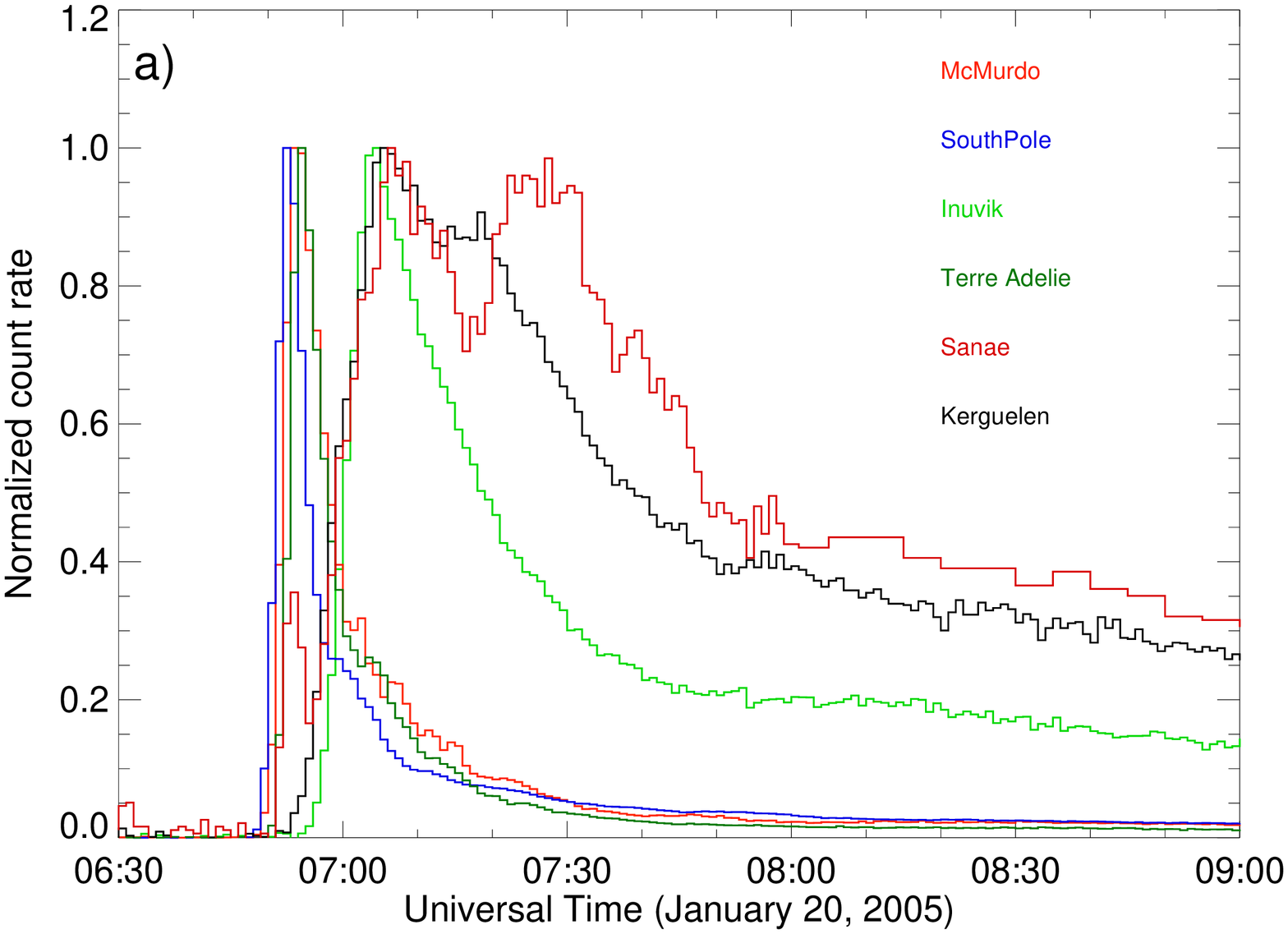}
   \hspace*{-0.011\textwidth}
	      }
\centerline{
   \includegraphics[width=0.41\textwidth,height=0.6\textheight,angle=90]{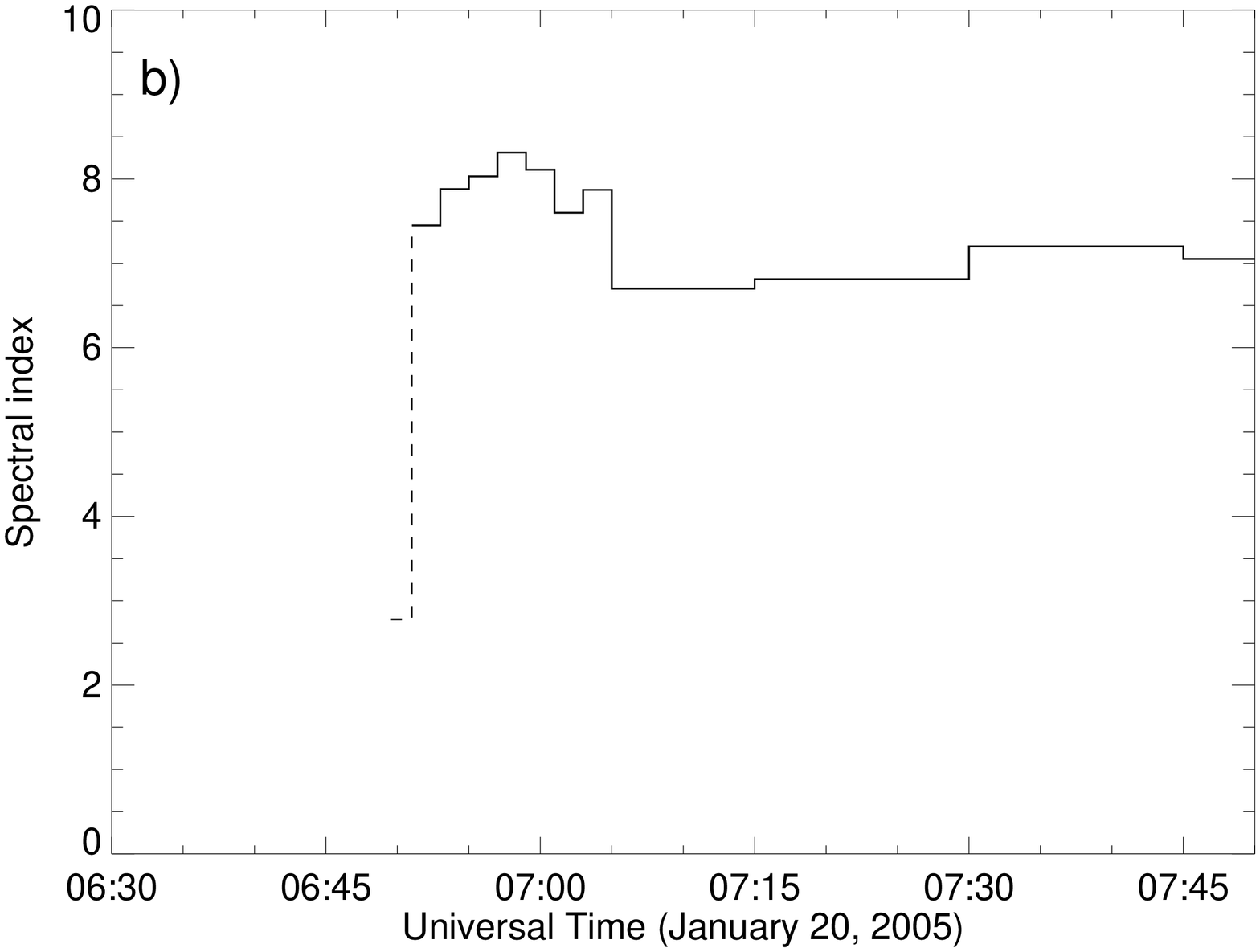}				}
  \centerline{	
  \includegraphics[width=0.41\textwidth,height=0.6\textheight,angle=90]{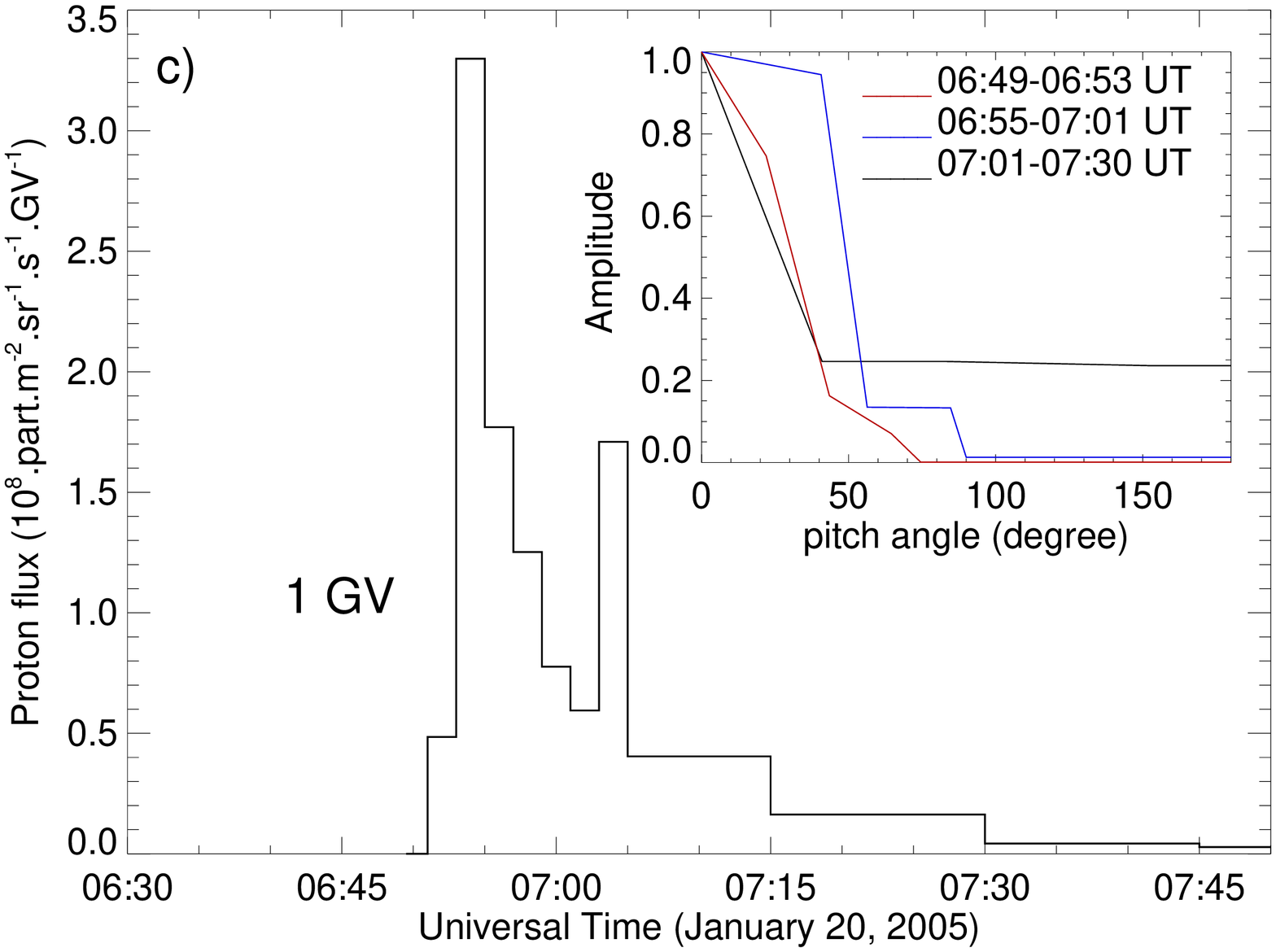}
}	
\centerline{	
  \includegraphics[width=0.41\textwidth,height=0.6\textheight,angle=90]{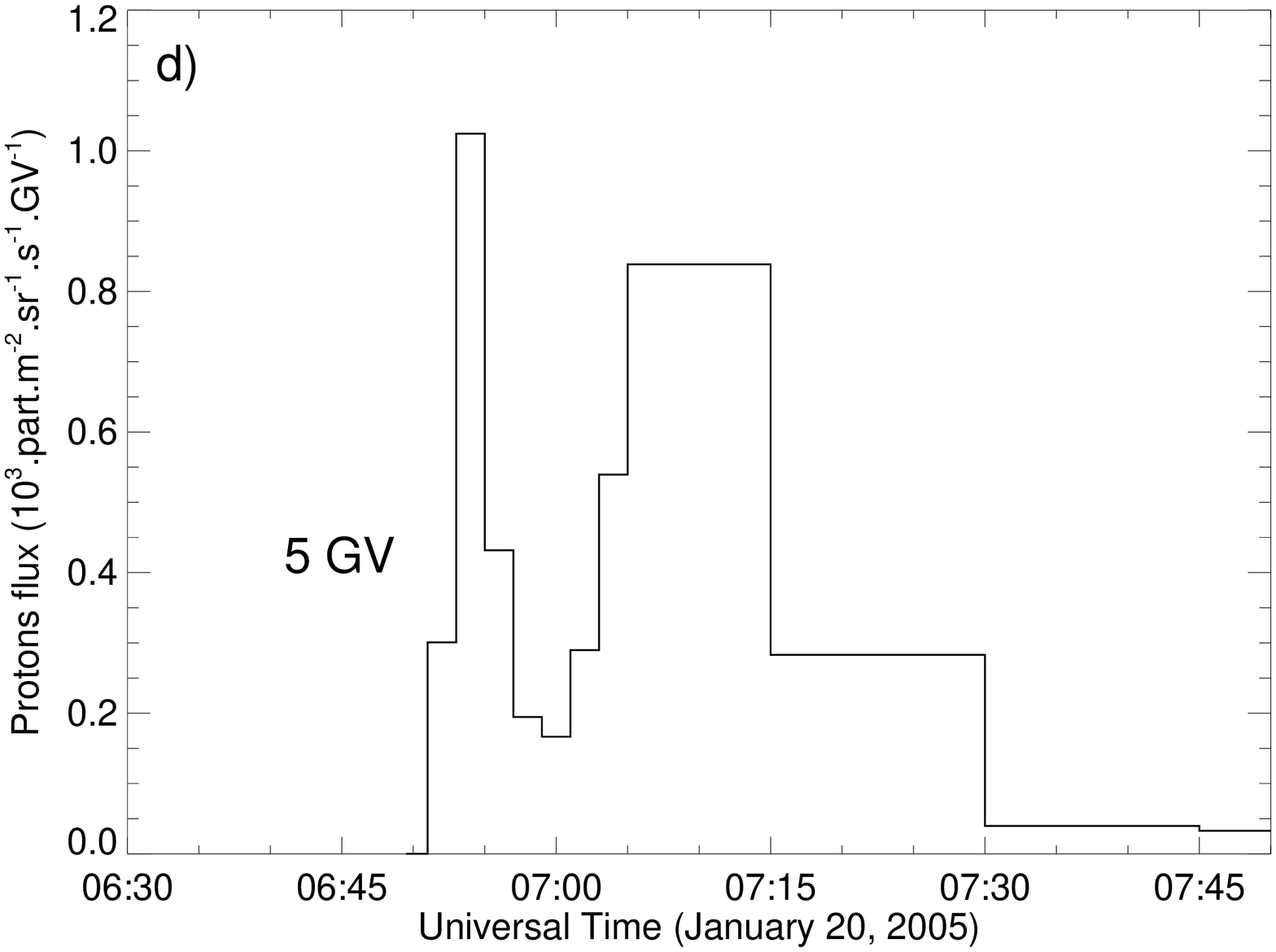}
              }       
\caption{
Time histories of observed and derived parameters of the relativistic particle event on 20 January 2005: {\rm a)} count rate time histories of several neutron monitors, the count rates are normalised by the individual maximum; {\rm b)} spectral index $\gamma(t)$ of the differential directional proton intensity spectrum ({\it cf.} Equation~(\ref{eq-flux})); {\rm c)} amplitude $A(t)$ (differential directional proton intensity at 1~GV rigidity) of the protons. Insert : pitch angle distributions; {\rm d)} differential directional proton intensity at $5$~GV rigidity. Note that the time scale in {\rm a)} is different from {\rm b)} to {\rm d)}.  
       }
\label{f-flux}
\end{figure}
 %

\subsection{Relativistic Protons at 1 AU}
\label{ss-NMs}

Neutron monitor time profiles of the GLE are shown in panel {\rm a)} of Figure~\ref{f-flux} and in several other papers \cite{Bie:al-05,Sim-06,Flu:al-05,Pln:al-07,Bom:al-08,McC:al-08}. Here we are interested in the profile of the primary particles outside the Earth's magnetosphere. The primary particles producing the GLE from the Sun are protons or occasionally also neutrons, but neutrons can be excluded in the present GLE. As seen in the top panel of Figure~\ref{f-flux}, the neutron monitors located in Antarctica (Sanae, McMurdo, South Pole and Terre Ad\'elie) start first and rise faster than the others. At these locations the zenith angle of the Sun is usually too large to detect solar neutrons by ground based cosmic ray detectors as neutron monitors or solar neutron telescopes, and there was virtually no difference of time profiles at high-altitude ({\it e.g.} South Pole) and sea-level ({\it e.g.} Terre Ad\'elie) stations. \inlinecite{Zhu:al-05} come to the conclusion that also the solar neutron telescope at Yangbajing (Tibet), at 4300~m above sea level, which observed the Sun at zenith angle $52^\circ$, did not detect any signature of solar neutrons.
 
The terrestrial magnetic field determines the viewing direction and the low-rigidity cutoff of a neutron monitor with respect to primary protons. A set of instruments distributed on Earth can therefore be used to determine the rigidity spectrum in a range between roughly 1 and 15~GV, and the angular distribution of the primary relativistic particles arriving at the top of the Earth's magnetosphere. The excess count rate of a neutron monitor is given by
\begin{eqnarray}
\Delta N(t)=\int_{P_{\rm c}}^{\infty}S(P)\cdot J_{\|}(P,t) \cdot F(\delta(P),t)\cdot dP
\label{eq-count}
\end{eqnarray}
where $P$ is the particle's magnetic rigidity, related to the total energy $E$ by $eP=\sqrt{E^2-m^2c^4}$. $e$ is the elementary charge, $P_{\rm c}$ is the low-rigidity cutoff imposed by the terrestrial magnetic field, $S(P)$ is the rigidity response of the neutron monitor, called yield function. $F(\delta(P),t)$ describes the angular distribution of the particle flux, where $\delta(P)$ is the angle between the direction of maximal solar particle flux and the asymptotic viewing direction of the neutron monitor at rigidity $P$. It describes the pitch angle distribution if the axis of symmetry is along the local magnetic field direction outside the geomagnetosphere. Based on the analysis of \inlinecite{Bue:al-06}, we assume in the following that this is indeed the case. $J_{\|}(P,t)$ is the rigidity dependent differential directional particle intensity (hereafter this differential directional proton intensity will be called ``proton intensity''),

\begin{eqnarray}
J_{\|}(P,t)=A(t) \left( \frac{P}{1~\rm {GV}} \right) ^{-\gamma(t)}. 
\label{eq-flux}
\end {eqnarray}

For the present study we employed the data of 40 neutron monitors of the worldwide network. The MAGNETOCOSMICS code developed by L.~Desorgher\footnote{http://reat.space.qinetiq.com/septimess/magcos/} has been used to simulate the propagation of energetic protons through the Earth's magnetic field and to determine for each neutron monitor the cutoff rigidity and the asymptotic directions from which the primary energetic particles impinged upon the magnetosphere. We computed for each neutron monitor the response to an anisotropic flux of charged particles at the top of the magnetosphere. Through a trial-and-error procedure that minimizes the difference between the modeled and the observed count rates for each monitor, we determined the amplitude $A(t)$, the spectral index $\gamma (t)$, and the anisotropy of incident particles at the boundary of the magnetosphere. We use in this method a variable time interval. At the beginning of a GLE the count rates and the spectrum change quickly. Later during the event the changes in the different parameters in time are less pronounced. In addition, the count rates of some neutron monitors showed a short pre-increase. Thus, we have chosen a time interval of $2$~min during the initial and the main phase of the GLE, and $10$~min and $15$~min during the recovery phase. More details can be found in \inlinecite{Bue:al-06}. 

Panel {\rm b)} of Figure~\ref{f-flux} displays the spectral index. The initial phase, where $\gamma=3$, is plotted by a dashed line, since we expect a hard spectrum of the solar cosmic ray flux near Earth in the first phase of the GLE just because the high energy particles will arrive at the Earth earlier than the protons with lower energies. 

Panels {\rm c)} and {\rm d)} display the time profiles of the proton intensities calculated by Equation~(\ref{eq-flux}) with $P=1$~GV ({\it i.e.} the amplitude $A(t)$) and $P=5$~GV. At the first peak of the impulsive proton profile intensity at 1 GV obtained from our analysis is $A=3.3\times10^8~\rm{protons~(m}^2\rm{~sr~s~GV)}^{-1}$, and the spectral index $\gamma = 8.0$. Using a similar method of analysis, \inlinecite{Bom:al-08} found $A=2.0\times10^8~\rm{protons~(m}^2\rm{~sr~s~GV)}^{-1}$ and $\gamma = 9.2$. \inlinecite{Pln:al-07} used a different description of the pitch angle distribution, but the index (7.6) and amplitude of the rigidity spectrum, when corrected for the different integration times, are similar to our values.

Several different approaches providing flatter spectra were reported in the literature. \inlinecite{Bie:al-05} and \inlinecite{McC:al-08} compared the signals observed with a traditional neutron monitor and with a monitor without lead shield, which is sensitive to lower energies than the shielded one. Their independent studies found spectral indices between 4 and 5 during the first peak of the GLE as observed, respectively, at the South Pole and Sanae stations. \inlinecite{Rya:Mil-05} derived $\gamma = 6.2$ from the Durham and Mount Washington monitors, which have different yield functions because they are located at different altitudes. Since the antarctic neutron monitors give information on the lower rigidities, and the north-american monitors on medium rigidities, we suggest that the different indices reflect a spectrum that is not a power law in rigidity, but gradually curves down to some high-rigidity cutoff, as has been shown by \inlinecite{Her:al-76}, \inlinecite{Lov:al-98} and \inlinecite{Vas:al-05}.

Hereafter we will use the proton intensity at rigidity $5$~GV, which corresponds to a kinetic energy of $4.15$~GeV, for the comparison with the electromagnetic emissions. Indeed, although the response function of neutron monitors depends on atmospheric depth, for most stations its maximum is close to $5$~GV \cite[their Figure~3]{Clm:Drm-00}. 
The time profiles of the relativistic protons display two well-identified peaks, which is a feature found in several other GLEs from the western solar hemisphere \cite{She:Sma-96,McC:al-08}. The short rise to maximum during the first peak requires an acceleration of particles to relativistic energies or their release within a few minutes at most. 

The inserted plot in panel {\rm c)} shows the pitch angle distribution for three time intervals between 06:49~UT and 07:30~UT. Between 06:49~UT and 07:01~UT, the first peak of the proton intensity time profile displays a narrow angular distribution. These protons suffer little scattering during interplanetary propagation. The maximum of the pitch angle distribution continues to be field-aligned during the second peak, but an increasingly broad range of pitch angles contributes. This suggests that particles came from the Sun throughout the first and the second peak of the proton time profile, but that interplanetary scattering \cite{Sai:al-05,McC:al-08} or a reflecting barrier outside 1~AU \cite{Bie:al-02,Sai:al-08} affected the second peak. 

The interplay between variations of the amplitude and the spectral index make the second peak appear much longer at $5$~GV than at lower rigidities. We consider this difference, which results from minor changes in the amplitude and spectral index, as questionable. In any case, the well-defined temporal structure of the proton intensity enables a more detailed comparison with coronal events than other GLEs.

%
%
\begin{figure}
\centerline{
   \includegraphics[width=0.9\textwidth,height=0.6\textheight,angle=90]{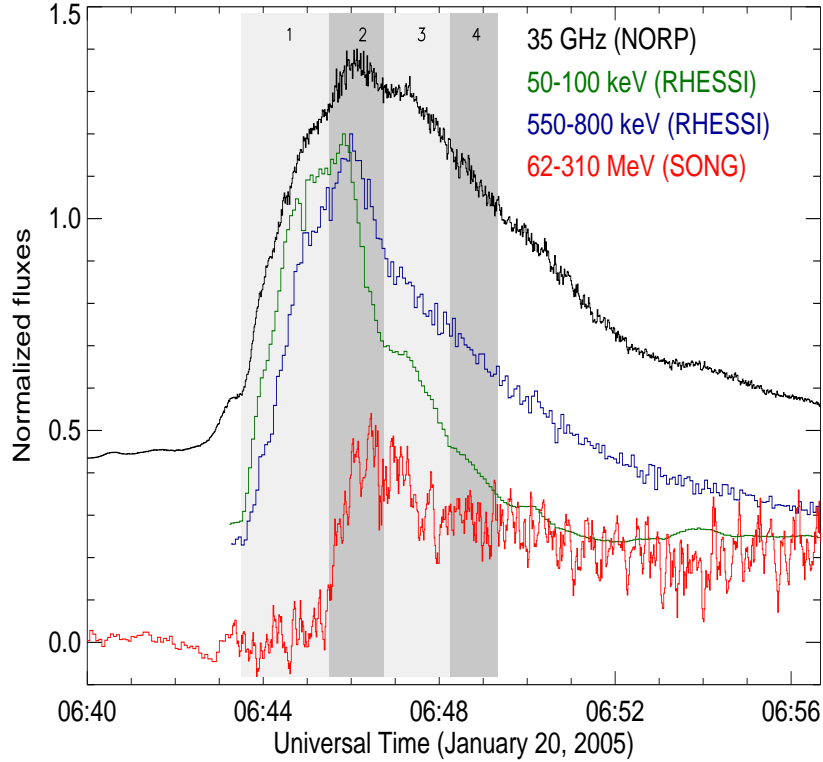}
   \hspace*{-0.011\textwidth}
               }       
\caption{
Time profiles of the normalised flux density at $35$~GHz (top) and normalised count rates of hard X-rays and $\gamma$-rays (RHESSI, CORONAS/SONG) at different energies. Each count rate is normalised by individual maximum and a term is added in order to separate properly the curves of each others. The $35$~GHz emission (Nobeyama Radio Polarimeter, courtesy K.~Shibasaki) is synchrotron radiation, emissions at ($50-100$) and ($550-800$)~keV are Bremsstrahlung. The high-energy $\gamma$-rays are pion-decay photons from primary protons at energies above $300$~MeV. Different episodes of particle acceleration are distinguished by vertical stripes numbered 1 to 4.
       }
\label{f-accph}
\end{figure}
%
%
%
%
%
%

%

\subsection{High-Energy Particles in the Low Corona and Chromosphere : $\gamma$-Ray, Hard X-Ray and Microwave Signatures} 
\label{ss-chrom}

High-energy electrons and protons in the low corona and chromosphere are revealed by their hard X-ray and $\gamma$-ray bremsstrahlung, gyrosynchrotron microwave emission, and different types of nuclear $\gamma$ radiation. Figure~\ref{f-accph} displays the time profiles observed by the Reuven Ramaty High Energy Spectroscopic Imager (RHESSI) \cite{Lin:al-02} in the photon energy ranges $50-100$~keV (green line) and $550-800$~keV (blue line), emitted by electrons with energies of order $100$~keV and $1$~MeV, respectively. The red curve shows CORONAS-F/SONG measurements \cite{Kuz:al-08} of $\gamma$-rays from $62-310$~MeV. 

At higher altitudes in the solar atmosphere, the relativistic electrons produced synchrotron emission in coronal loops. The Nobeyama Radio Polarimeter \cite{Nak:al-85} measures whole Sun integrated flux densities at selected frequencies in the band $1$ to $80$~GHz. In Figure~\ref{f-accph}, we plot only the time profile at $35$~GHz (black curve), since the emission is self absorbed at the lower frequencies. A synchrotron spectrum is cut off at the characteristic frequency $\nu = 3/2 \gamma^2 \nu_{\rm {ce}}$, where $\nu_{\rm{ce}}$ is the electron cyclotron frequency and $\gamma$ is the Lorentz factor. If the magnetic field is $500$~G, $35$~GHz emission is hence emitted by electrons with kinetic energy $1.5$~MeV.

These four time profiles have common structures that reveal distinct episodes of particle acceleration during the flare. We defined four main episodes of particle acceleration corresponding to time intervals with a distinct peak in one or several of these spectral ranges. These episodes are highlighted in Figure~\ref{f-accph} by different tones of grey shading, and are labelled from 1 to 4. Other distinct rises of emission produced by energetic electrons occur later, {\it e.g.} near 06:53~UT.

The hard X-ray and microwave emissions start to rise before 06:44~UT (between 06:42~UT and 06:43~UT at $35$~GHz) and display several peaks. The time profiles of hard X-rays present one peak in each of the acceleration phases 1 and 2. The rise to the second peak hides the decrease from the first and vice versa. Both peaks are also seen in the $35$~GHz time profile. The initial rise of the hard X-rays is faster, and the first peak is more pronounced, in the $50-100$~keV range than in the $550-800$~keV range. Hence relatively more high energy electrons are accelerated during the second acceleration episode than during the first. This reflects the continued spectral hardening throughout most of the event reported by \inlinecite{Sld:al-08}. 
%
%
\begin{figure}
\centerline{
    \includegraphics[width=0.5\textwidth,height=0.6\textheight,angle=90]{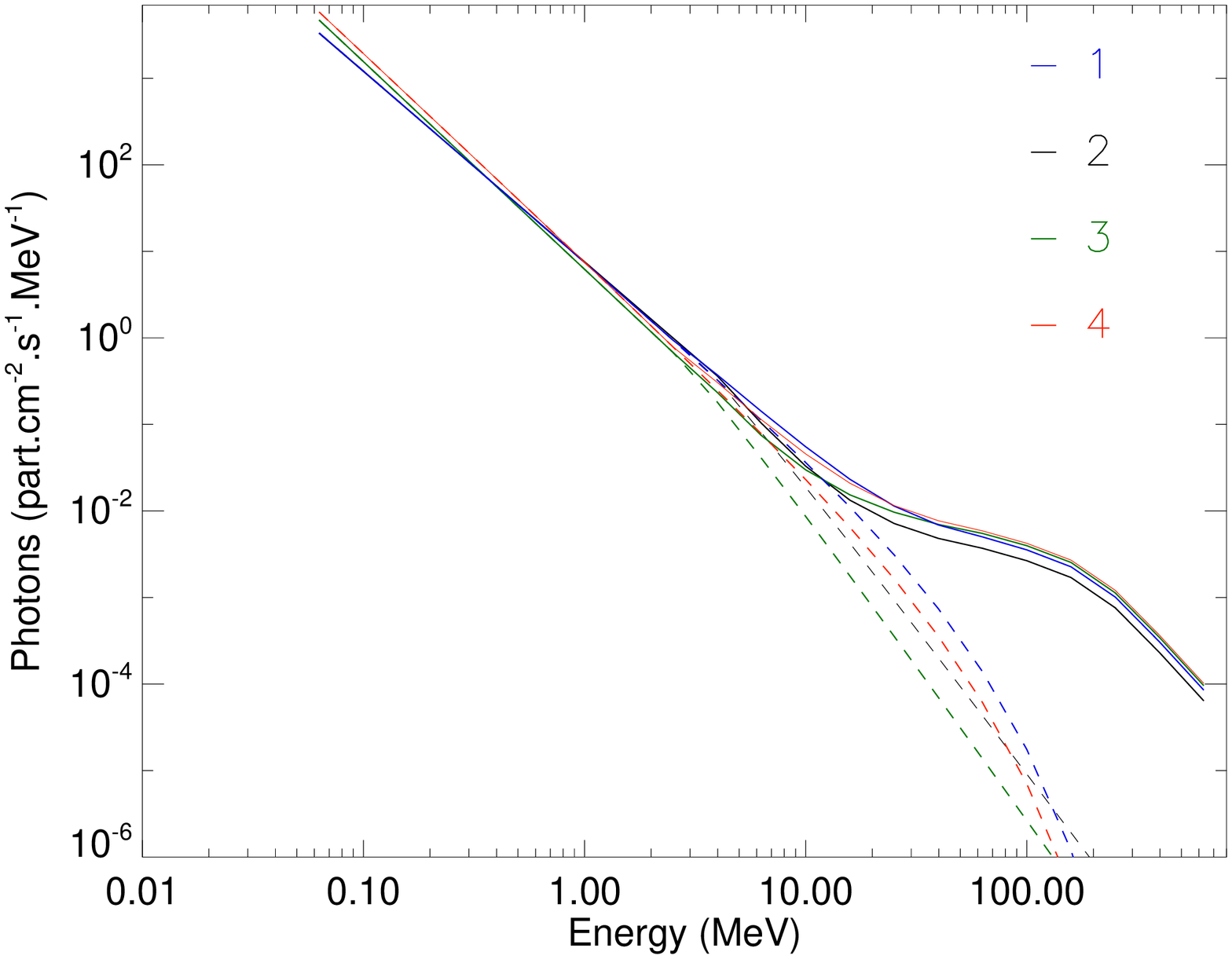}
}	
\centerline{
  \includegraphics[width=0.5\textwidth,height=0.6\textheight,angle=90]{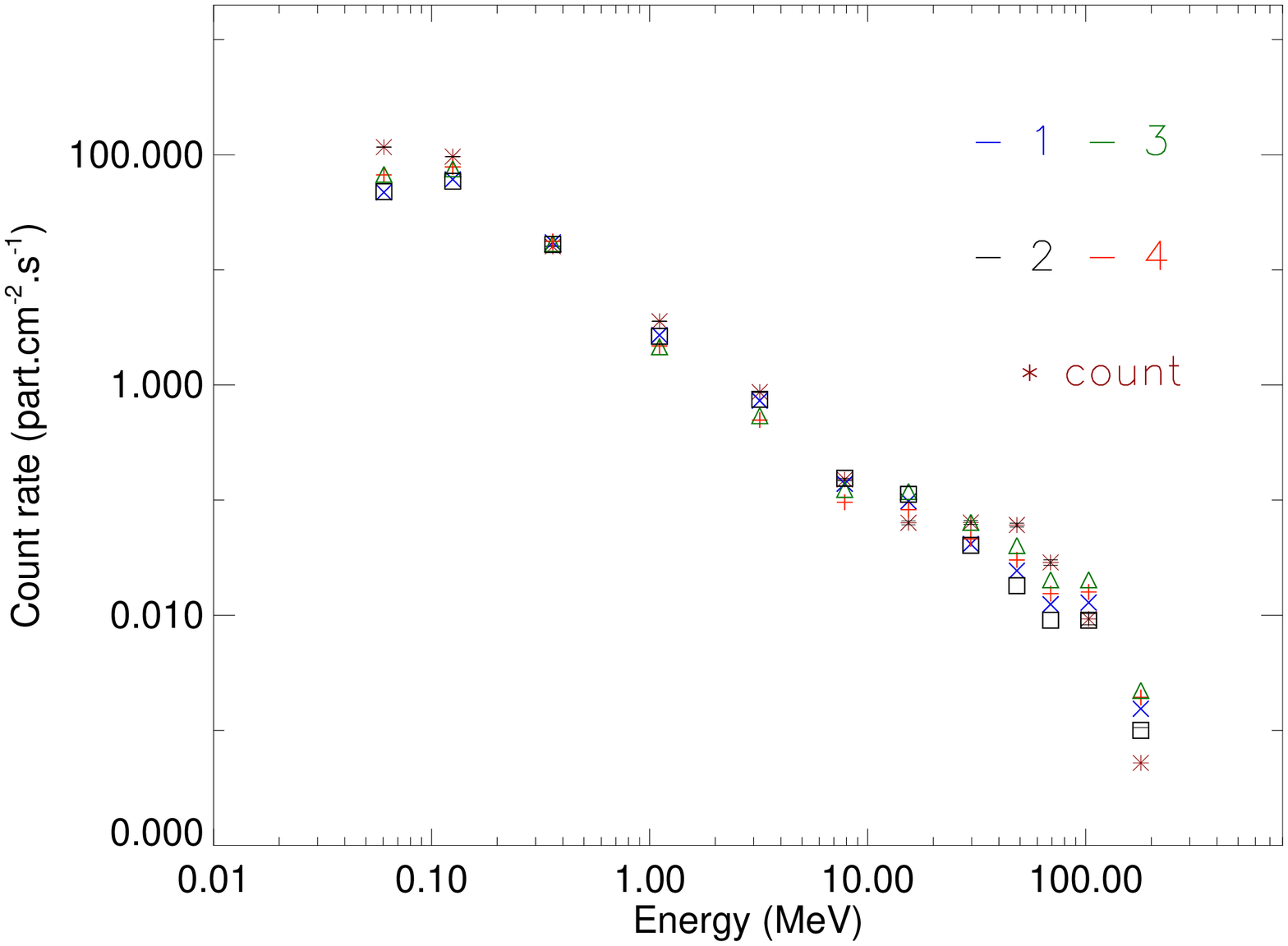}
               }    
\centerline{
    \includegraphics[width=0.5\textwidth,height=0.6\textheight,angle=90]{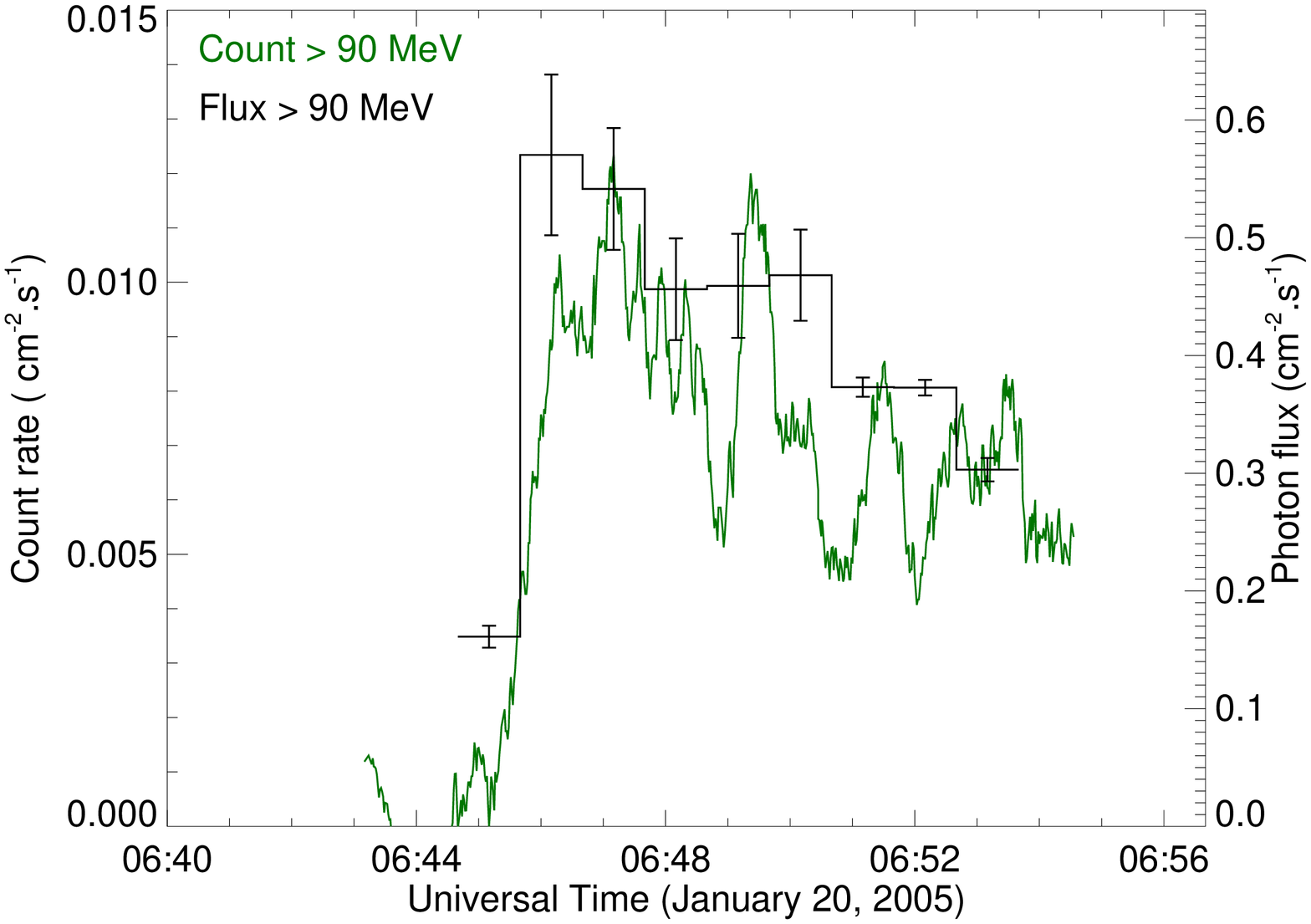}
}
\caption{
High-energy photon emission observed by CORONAS/SONG: The top panel displays the model photon spectra with an electron bremsstrahlung and a pion decay component [Equation~(\ref{eq-spec})]. Solid lines represent total spectra, dashed lines the electron bremsstrahlung components for power laws with indices 2.2 and 2.4 and an exponential rollover (curves numbered 1 and 4; {\it cf.} Equation~(\ref{eq-RO}); $\epsilon_0=30$~MeV) or a double power law (curves 2 and 3; {\it cf.} Equation~(\ref{eq-DP}); $\epsilon_{\rm {cut}}=30$~MeV). On the middle panel, the observed count rate spectrum and fits with the same coding colors than the model. On the bottom panel, the count rate (30~s smoothed) and photon flux at photon energies above 90~MeV.
	  }
\label{f-SONG}
\end{figure}
%

The appearance of increasingly high particle energies during episode 2 is clearly shown by the $\gamma$-rays above 60~MeV (Figure~\ref{f-accph}, bottom). The count rate starts to rise between 06:43 and 06:45~UT, consistent with an early start of acceleration phase 2, which is hidden in the electron radiation profiles by episode~1. We use as the earliest time the start of the high-energy $\gamma$-ray emission at 06:45:30~UT. 

The flux of $\gamma$-rays above 60~MeV can be due to bremsstrahlung emission of energetic electrons or to the decay of neutral pions produced by protons above $300$~MeV. To separate the two components, we fitted the response of the SONG detector to a combined bremsstrahlung and pion decay spectrum 
\begin{equation}
F(\epsilon) = w_{\rm b} (\epsilon)^{-\gamma} C(\epsilon) + w _\pi \Phi(\epsilon)
\label{eq-spec}
\end{equation}
($\epsilon$ is the photon energy). The bremsstrahlung spectrum is the combination of a power law with a function describing the steepening at high photon energies. The steepening is either represented by a spectral rollover

\begin{equation}
C(\epsilon) = \exp(-\epsilon /  \epsilon_0 ) 
\label{eq-RO}
\end{equation}
or by a function that describes a gradual transition between power laws with index $\gamma$ at low energies and $\beta \ge \gamma$ at high energies: 

\begin{equation}
C(\epsilon)=1 \; {\rm if} \; \epsilon \leq \epsilon_{\rm {cut}} 
\end{equation}

\begin{equation}
C(\epsilon) = [ 
	\frac{1+\epsilon/ \epsilon_0}{1+\epsilon_{\rm {cut}}/\epsilon_0} ]^{(\gamma-\beta)} 
	\; {\rm if} \;  \epsilon > \epsilon_{\rm {cut}} .
\label{eq-DP}
\end{equation}

Nuclear lines were neglected in the fitting procedure, because the count rate in even the strongest one (at $2.2$~MeV) does not exceed $15$\% of the underlying continuum. 

\begin{figure}
\centerline{
   \includegraphics[width=0.9\textwidth,height=0.9\textheight,clip=]{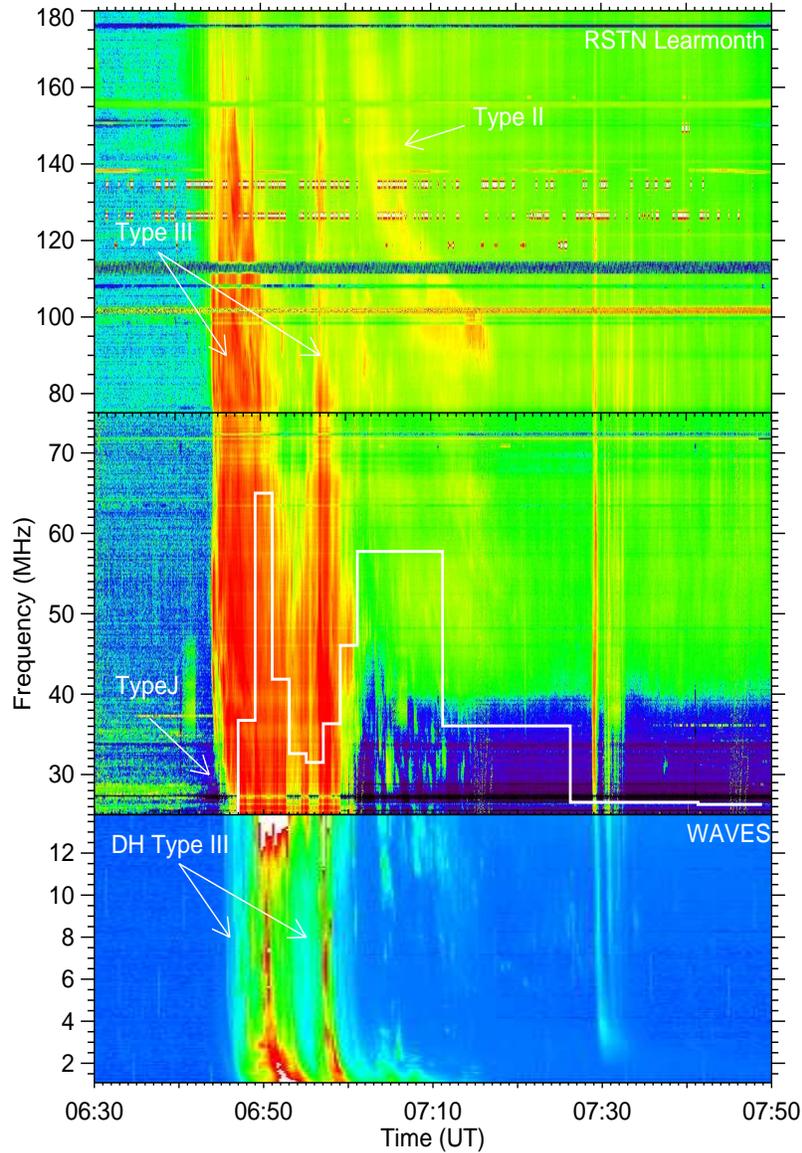}
   \hspace*{-0.011\textwidth}
            }       
\caption{
Dynamic spectrum of the metric-to-decametric radio emission during the 20 January 2005 event, observed at the Learmonth station of the RSTN network ($180-25$~MHz) and the WAVES spectrograph aboard the {\it WIND} spacecraft ($14-1$~MHz). The overplotted white curve is the time history of the proton intensity at $5$~GV rigidity (kinetic energy $4.15$~GeV), shifted backward by $216$~s (see Section~\ref{s-ip} for the evaluation of the time shift).
        }
\label{f-spec}
\end{figure}

The spectral shape of the pion decay component was calculated by R.~Murphy (private communication, 2007) and is similar to other commonly used forms \cite{Mur:al-87}. The pions are considered to be produced by particles (protons and 50\% $\alpha$ particles) with a power law spectrum that extends up to 1.5~GeV and an isotropic angular distribution in the downward direction. The calculations demonstrate that the photon intensity and the location of the spectral maximum in the energy range up to 300 MeV are identical for power law proton spectra with index 2 and 3. Because the highest energy interval of SONG is $150-300$~ MeV, we use the pion decay component calculated by R.~Murphy with the proton index 3. The detector response to a parallel beam of $\gamma$-rays is computed using GEANT3.21 routines, and the spectral parameters are derived by minimising ($\chi ^2$) the difference between observed and modelled count rates in selected detector channels. A set of model spectra and the resulting fit to the observed count rate spectrum during the early phase of the high-energy photon emission are plotted in Figure~\ref{f-SONG}. The count rate at photon energies above $90$~MeV, smoothed over 30~s, and the flux, which at those photon energies is largely dominated by pion decay photons are also shown. We conclude that the high-energy photon profile in Figure~\ref{f-spec} and in Figure~\ref{f-accph} can be attributed to high-energy protons in the solar atmosphere, requiring $2.3\times10^{31}$ protons above $30$~MeV during a $1$~min interval at the peak of the emission. For comparison, \inlinecite{Vil:al-03a} evaluated in their Table~3 that five to ten times more protons were needed for the pion decay $\gamma$-ray emission during the 24 May 1990 event.

We can make a rough comparison of the number of protons detected during the GLE and the number of protons required for the pion decay $\gamma$-ray emission. The pitch angle distribution (insert of Figure~\ref{f-flux}) shows that all protons during the first peak stream anti-sunward. If we assume that they are injected at the Sun into a flux tube with diametre $3 \times 10^9$~m at $1$~AU (corresponding to an angle of $1.2^{\circ}$), as derived for impulsive electron \inlinecite{But-98} and ion events \inlinecite{Maz:al-00}, our measured proton intensity at the peak of the GLE implies that $6 \times 10^{28}$ with rigidities above $1$~GV are released to interplanetary space during the $2$~min interval around the maximum of the GLE. The number increases to $3.5 \times 10^{31}$ if a greater angular range of 30$^\circ$, corresponding to a size of $7.8 \times 10^{10}$~m at $1$~AU, is used. This result is to be considered with caution since we use a (steep) power-law as rigidity spectrum. As discussed earlier, the real spectrum is expected to consist of a flatter power law at low rigidities with a steep falloff or cutoff above some limiting rigidity. Thus, if this limiting rigidity is near or above $1$~GV, our evaluation of the proton intensity above 1~GV with the power law will be overestimated. For comparison, the spectrum that generates the pion decay $\gamma$-ray emission contains about $10^{29}$~protons above $1$~GV (kinetic energy $433$~MeV), which is within the broad range of values estimated for the escaping protons.

In summary, the energetic particle signatures in the low atmosphere during the impulsive flare phase reveal several successive episodes of particle acceleration with durations of the order of a minute. The most energetic one starts close to 06:45:30~UT, judging from the $\gamma$-rays and the accompanying peaks in the hard X-ray and microwave emissions of energetic to relativistic electrons. The most energetic protons and electrons hence start to be accelerated during this episode, more than 2~min after the first radiative signatures of high-energy electrons in the solar atmosphere.

%
%
%

\subsection{Radio Evidence on Particle Escape to the High Corona and Interplanetary Space}
\label{ss-wind} 

Metric-to-hectometric radio emissions of non-thermal electrons from the middle corona to the interplanetary medium are measured respectively by the RSTN network\footnote{http://www.ngdc.noaa.gov/stp/SOLAR/ftpsolarradio.html\#spectralgraphs} ($25-180$~MHz; we use data from the Learmonth station) and the WIND/WAVES experiment in the range 0.04-14~MHz \cite{Bou:al-95}. The combined spectrum of the two instruments in the $1-180$~MHz band, corresponding roughly to a range of heliocentric distances between $1.2$ and $10-20$~R$_\odot$, is represented in Figure~\ref{f-spec}. Different burst groups can be distinguished.

\begin{figure}[ht]
\centerline{
\includegraphics[width=0.9\textwidth,height=0.6\textheight,angle=90]{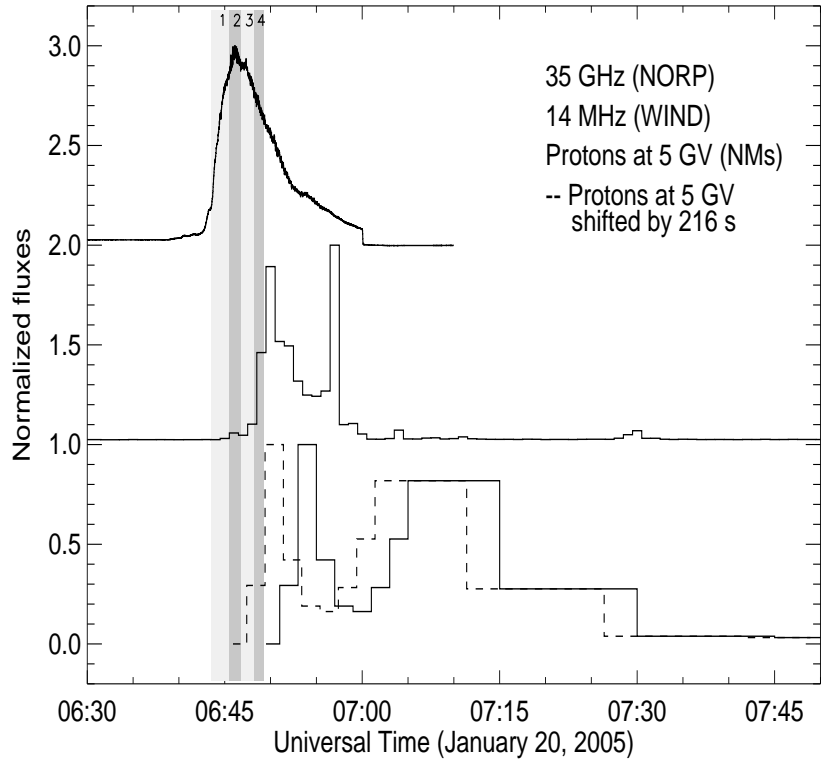}
   \hspace*{-0.011\textwidth}
               }       
\caption{
Temporal evolution of the normalised flux densities in microwaves, {\it i.e.} gyrosynchrotron emission of relativistic electrons (top; $35$~GHz; Nobeyama Radio Polarimeter), decametric waves (middle; $14$~MHz, WIND/WAVES radio spectrograph) emitted by unstable electron beams in the high corona, the normalised proton intensity at Earth ($5$~GV rigidity, kinetic energy $4.15$~GeV; solid line), and the normalized proton intensity shifted by $-216$~s (dashed line). Each flux is normalised by individual maximum and a term is added in order to separate properly the curves of each others. The time axis refers to the time of detection at the WIND spacecraft or at Earth.
 }
\label{f-mplt}
\end{figure}

Two groups of decametric-to-hectometric (henceforth DH) type~III bursts are seen in the WIND/WAVES spectrum between 1 and 14~MHz, respectively from 06:45~UT to 06:55~UT and 06:57~UT to 07:00~UT. Using the individual time profiles, we find that the first group starts with a faint burst at 06:45~UT $\pm30$~s, followed by bright bursts starting at 06:46~UT$ \pm30$~s (see the time profile at $14$~MHz, the central curve in Figure~\ref{f-mplt}). 

The two bright DH type~III groups extend up to metre waves seen by RSTN. The starting frequency is above the RSTN high-frequency limit of $180$~MHz. With the usual assumption that metric type~III bursts are emitted at the harmonic of the electron plasma frequency, this implies that the acceleration sites are in the low corona, where the ambient density is above $10^8$~cm$^{-3}$. The first metre wave burst in the RSTN spectrum, however, does not seem to continue into the WAVES band, but bends around to form a type~J burst between 35 and 25~MHz. This is probably not an artefact of decreasing sensitivity near the low-frequency border of the receiver, because no comparable decrease is observed in the subsequent emissions. Hence electrons accelerated during the first acceleration episode do not seem to get access to interplanetary space, but propagate in closed coronal magnetic structures.

Thus, we assume that the first injection of non-thermal electron beams into interplanetary space is traced by the first faint DH type~III burst at 06:45~UT$\pm30$~s, and is followed by a series of more intense injections corresponding to the first bright type~III group.

Shortly after the second type~III group, the RSTN spectrum shows a slowly drifting band of less intense emission. It is observed between $85$~MHz and $180$~MHz from 07:00~UT to 07:15~UT. The frequency drift rate is $0.1-0.2~\rm{MHz~s}^{-1}$, and the instantaneous bandwidth about 30\% of the low frequency limit of the burst. This is typical of a type II burst at long metre wavelengths \cite{Man:al-95}. The burst therefore shows the propagation of a shock wave through the corona. If it is harmonic plasma emission, the ambient electron densities in the metric type~II source range from $10^8$ to $2 \times 10^7~\rm{cm}^{-3}$. At lower frequencies the WIND/WAVES spectrum presents a weak type II emission between 07:03~UT and 07:30~UT, without any harmonic relationship to the metric type~II burst.

Underneath the structured type~III and type~II bursts the dynamic metre wave spectrum shows a diffuse background (green in Figure~\ref{f-spec}) that reveals gyrosynchrotron emission from mildly relativistic electrons. It starts together with the first metric type~III bursts and continues throughout the plotted time interval. It can be identified until at least 07:50~UT in single frequency records, with broadband fluctuations from centimetre to metre waves that suggest repeated electron injections.

\section{Relativistic Protons at the Earth, Coronal Acceleration, and Interplanetary Propagation}
\label{s-ip}

%
%
%

\subsection{Interplanetary Propagation and Initial Solar Release Time}
\label{ss-obs}

The solar wind speed measured by WIND/SWE \cite{Ogi:al-95} at the time of the GLE is 800~km~s$^{-1}$, implying a Parker spiral of length $1.05$~AU. The relativistic protons at 5~GV (corresponding to a velocity of $0.98c$) are then expected to arrive $46$~s after the electromagnetic emission. But the first relativistic protons were detected not earlier than 06:49~UT \cite[their Figure 6]{McC:al-08}, 3 to 6 minutes after the start of the hard X-ray, microwave, and $\gamma$ radiations in the solar atmosphere.

However, since the relativistic protons were observed at Earth during a Forbush decrease, it is clear that the interplanetary magnetic field was far from nominal during this event. This is confirmed by the actual magnetic field measurements \cite{Pln:al-07,McC:al-08}. We therefore compare the time profiles of the proton intensity at $1$~AU with the radio emissions in the corona to attempt a more realistic evaluation of the interplanetary path length and eventually of the solar release time of the particles detected near and at the Earth.

In Section~\ref{ss-wind}, we concluded that the first DH type III bursts ($14$~MHz) reveal the first injection of non-thermal electron beams in the interplanetary space. The time profiles of the radio emission at $14$~MHz and of the proton intensity at $5$~GV are shown in Figure~\ref{f-mplt}. We note that both time profiles display similarly rapid initial rise phases and, broadly speaking, two peaks. These similarities suggest a common release of the relativistic protons and the radio emitting electron beams. 

 The rise phases and the first peak of the time profiles of protons and the bright type~III emission at 14~MHz coincide when the proton profile is shifted backward by $t_{\rm {shift}}=216$~seconds. The time shift $t_{\rm {shift}}$ between the two profiles corresponds to the supplementary path length travelled by the protons. Figure~\ref{f-mplt} displays the original (black solid line) and the backward shifted time profile (black dashed line) of the relativistic protons at $5$~GV. The timing would not have been changed significantly if we had used a time profile at lower rigidity. Given the velocity of $0.98c$ and the light travel time of 489~s from Sun to Earth, the delay of $216$~s implies that the protons travelled a distance of about $1.38$~AU in interplanetary space. 

The delay of $216$~s gives only a lower limit of the travel time, because the acceleration region is presumably much closer to the Sun than the $14$~MHz source. We determine the upper limit of the path length, assuming that the first relativistic protons are injected immediately after the acceleration of the protons that created pions in the low solar atmosphere. Thus we evaluate as an upper limit of the supplementary travel time of the protons with respect to photons $t_{\rm {shift}}=4.5$~min, which induces a path length between the acceleration site of the first relativistic protons and the Earth of $0.98c \times (489$~s$ + t_{\rm {shift}})=1.49$~AU.

The time interval between the onset of the second acceleration episode and the bright $14$~MHz emission of electron beams (06:46~UT) is of $30$~s. With the density determination of \inlinecite{Kou-94}, the $14$~MHz source is located about $4~R_{\odot}$ above the photosphere. Thus the electrons injected in the low corona together with the relativistic protons have to travel $ 4~R_{\odot}$ in $30$~s. This implies an electron beam velocity of roughly $0.3c$, which is a classical velocity for electron beams producing type III bursts in the corona \cite{Suz:Dul-85,Poq:al-96}. \inlinecite{McC:al-08} evaluate a longer path length of $1.76$~AU. They also assume that the injection of relativistic protons is simultaneous with the beginning of the high-energy $\gamma$-ray emission. The difference of $0.27$~AU is mostly due to the definition of the onset time. These authors use a strongly smoothed version of the $\gamma$-ray profile, from which they deduce an onset time between 06:43 and 06:44~UT. In our Figure~\ref{f-accph}, this time corresponds to the beginning of the first acceleration episode. But at this time pion-decay gamma emission has not yet started, as shown in Figure~\ref{f-accph} (see also Figure~4 of \inlinecite{Kuz:al-06}).

As an alternative, we also checked the frequently used assumption that all energetic particles accelerated during a flare start to be released together with the onset of the electromagnetic emission. If we assume that the escaping protons were accelerated and released with the first microwave emitting electrons, their travel time would be $489$~s$+t_{\rm {shift}}=489 +506$~seconds, and the interplanetary path length $1.95$~AU. While this value is not inconsistent with particles travelling along a twisted magnetic field line within an interplanetary CME \cite{Lar:al-97}, it is in conflict with the idea suggested by the DH type III emission that the first microwave and hard X-ray emitting electrons have no access to interplanetary space. It is of course also inconsistent with the timing of the pion decay gamma ray emission, which starts several minutes after the first hard X-ray and microwave signatures.

\subsection{Release of Relativistic Protons and Associated Solar Activity Throughout the Event}
\label{ss-time}

Figure~\ref{f-spec} shows the superposition onto the radio spectrum of the $5$~GV proton profile in white, shifted backward in time by $216$~s so as to correct the travel path difference between protons and photons.

By construction the first peak in the proton time profile starts with the first group of DH type III bursts. Moreover its duration is comparable with that of the DH type~III group (about $10$~min). Thus we conclude that the proton intensity time profile at the Earth and the radio time profile of the first DH type~III group in the high corona are similar and that the first protons are injected in the interplanetary space with the first escaping non-thermal electrons. This injection time coincides with the second episode of impulsive coronal particle acceleration, hence with the start of the $\gamma$-ray emission due to protons above 300~MeV in the solar atmosphere, and with a fresh electron injection with harder spectrum than before. Particles accelerated during and after the second acceleration episode get direct access to interplanetary space. 

The time profile of the proton intensity shows a second peak that lasts longer than the first. Provided the path length of the relativistic protons is the same as during the first peak, the onset of this second peak coincides with the second group of type III bursts seen from $1$ to $180$~MHz, and the prolonged tail of the peak is accompanied by the type II burst between $75$ and $180$~MHz.

As noted by \inlinecite{Poh:al-07}, the shock wave producing this type II emission is not the bow shock of the CME observed by SOHO/LASCO. Indeed, at 06:54 UT, a few minutes before the type~II burst, the CME front is at heliocentric distance 4.5~R$_\odot$, where electron densities inferred from eclipse measurements are below $2 \times 10^6$~cm$^{-3}$ \cite{Kou-94}, corresponding to plasma frequencies below $13$~MHz, much lower than those of the metric type~II burst. 

The exciter speed of the type~II burst is also very different from the CME speed : the measured relative frequency drift rate is $-6.8 \times 10^{-4}$~s$^{-1}$. Such a drift is produced by an exciter that moves at a speed of roughly 500~km~s$^{-1}$ along a hydrostatic density gradient (electron-proton plasma, $T=1.5 \times 10^6$~K) at heliocentric distance 2~R$_\odot$. This is much lower than the speed of the CME, which \inlinecite{Grc:al-08} estimated between 2000 and 2600~km~s$^{-1}$.

\section{Discussion}
\label{s-disc} 

\subsection{Summary of the Observational Findings}
\label{ss-sum}

From the preceding analysis we infer the following scenario:

\begin{itemize}

\item The particle acceleration in the corona deduced from the electromagnetic signatures (Section~\ref {ss-chrom}) has several episodes in the impulsive phase, with individual durations of the order of $1$~min.

\item Protons with energies above $300$~MeV start to be accelerated during the second acceleration episode within the impulsive phase, together with electrons that have a harder spectrum than during the first episode (Section~\ref {ss-chrom}). 

\item The electrons accelerated during the first episode remain confined in closed coronal magnetic structures, while particles accelerated during the second and later episodes have access to the interplanetary space along open magnetic structures (Section~\ref {ss-wind}).

\item The first rise of the relativistic proton profile at Earth is due to protons that are accelerated during the second episode of the impulsive phase.

\item A second rise of the relativistic proton profile occurs after the peak of the hard X-ray and microwave emission. The onset of this second peak coincides with a new acceleration of electrons in the low corona during the decay of the microwave and hard X-ray burst, and with a fresh injection of electron beams into interplanetary space (second group of type III burst seen on Figure~\ref{f-spec}). It is accompanied by shock-related radio emission in the corona at heliocentric distances below 2.5~$R_\odot$.

\end{itemize}

\subsection{Evidence for Relativistic Proton Acceleration in the Flaring Active Region}
\label{ss-disc1}

Relativistic protons arrive at Earth during this GLE with some delay with respect to the first radiative signatures of particle acceleration and the travel time along the nominal Parker spiral. The delay comes on the one hand from the interplanetary path length, which is longer than the nominal Parker spiral. But a key for understanding the timing of particle acceleration during this event is the identification of elementary episodes within the impulsive flare phase. The first peak of the relativistic proton time profile is related to the impulsive phase of the flare, but to the second identified episode of particle acceleration. Similarly to the escaping relativistic protons, the bulk of the high-energy protons producing indirectly high-energy gamma rays through neutral pions are also accelerated in the second episode. The close connection between the interacting and escaping high-energy protons is corroborated by the finding that the number of protons required for the GLE would also be sufficient to produce a pion decay $\gamma$-ray excess.

Release delays of GLE protons with respect to the first electromagnetic signatures are well etablished \cite{Car-62,Cli:al-82}. They are often attributed to the trapping of the accelerated particles in closed magnetic structures or to the acceleration in a shock wave at greater altitude than the flaring loops \cite{Loc:al-90,Kah-94}. But the delay observed on 20 January 2005, among the shortest ever found, is consistent with a distinct acceleration episode in the impulsive flare phase. Such delays have been regularly reported for pion-decay $\gamma$-ray emission in other flares \cite{Frr:al-86,Deb:al-97,Dnp:al-99,Tro:al-08}, and had been identified in the SONG data of the 20 January 2005 flare by \inlinecite{Kuz:al-08} and \inlinecite{Grc:al-08}. Acceleration delays of relativistic protons on time scales of a few minutes hence seem to be a general feature of large flares. 

Imaging observations of this event at hard X-rays and $\gamma$-rays \cite{Kru:al-08} show the usual configuration of bright chromospheric footpoints of coronal loops, on top of UV ribbons, together with a presumably coronal $\gamma$-ray source. Such observations are commonly ascribed to a complex magnetic topology implying magnetic reconnection in the low corona. So the timing, the energetics, and the X/$\gamma$-ray source configuration during the 20 January 2005 event are consistent with a common origin of the interacting protons and the protons producing the first GLE peak in the flaring active region.

\subsection{The Second Peak of the Relativistic Proton Profile}
\label{ss-peak2}

The correspondence between radio emissions and the second peak in the proton time profile is more complex than for the first peak.

The duration of this second peak and its occurrence in the decay phase of the hard X-ray and microwave emissions suggest a difference in the acceleration region, the conditions of particle propagation, or both. The finding that this second peak is accompanied by distinct metric radio emission and by some weak microwave and hard X-ray signature can, however, be considered as a further argument that the peak is due to a fresh injection of protons at the Sun, rather than being due to changed conditions of interplanetary propagation as argued by \inlinecite{Sai:al-05}. The metric radio observations suggest two possible scenarios :

The first builds on the association of the start of the proton rise with a new, short group of type~III bursts and a faint rise of the decaying microwave and hard X-ray time profiles. If the relativistic protons are injected simultaneously with the non-thermal electron beams, during about 3~min (photon arrival time 06:57~UT - 07:00~UT), the prolonged tail would be ascribed to interplanetary scattering or reflection beyond 1~AU. Scattering requires very different propagation conditions from the first proton peak, which \inlinecite{Sai:al-05} attribute to perturbations created by the first proton beam. The prolonged presence of gyrosynchrotron radiation in the corona could also argue for prolonged particle acceleration there, without any need to invoke interplanetary transport for the duration of the relativistic proton release. But we have no common feature in the timing to corroborate this idea.

An alternative scenario is based on the association with a metric type~II burst during the maximum of the second relativistic proton peak and its decay phase, which may suggest shock acceleration of the second peak as proposed for different reasons by \inlinecite{McC:al-08}. \inlinecite{Poh:al-07} extrapolate the type~II radio band backward to a start near 06:54~UT at 600~MHz (their Figure~13). If this is correct, the entire second peak of the proton profile is accompanied by the radio signature of a coronal shock. This shock is seen at an altitude below 2.5~$R_\odot$ and at least its radio emission, if not the shock itself, is of short duration. Both findings are more consistent with a lateral shock than with a radially outward driven shock in front of the CME. A lateral shock would be expected to be quasi-perpendicular with respect to open magnetic flux tubes in the low corona (see Figure 2 of \citeauthor{Vai:Kha-04}, \citeyear{Vai:Kha-04}). The initially lateral shock would become quasi-parallel to the open field lines upon propagating outward. It has been argued elsewhere \cite{Tyl:Lee-06} that quasi-parallel shocks are less efficient particle accelerators at high energies than quasi-perpendicular shocks. This could explain the short duration of this type II burst and the duration of the second proton peak.

\subsection{Multiple acceleration episodes in solar energetic particle events}

GLE scenarios including two components, called a prompt one and a delayed one, had been introduced before \cite{Tor:al-96,Mir:al-00}. The delayed component was ascribed to acceleration at a CME-driven shock wave. While the present analysis contains one scenario that is consistent with this two-component injection, it also shows that the coronal acceleration history is much more complex: there is no unique flare-related acceleration, but the impulsive flare phase is itself structured, as has long been known from hard X-ray observations \cite{deJ:deJ-78}. If a coronal shock wave accelerates relativistic protons in a later phase of the event, it is not necessarily the bow shock of the CME which is the key element. Clearly, detailed comparative timing analyses of GLEs and flare/CME tracers provide relevant constraints to understand the origin of relativistic particles at the Sun.

\begin{acks}
 This research was supported by the French Polar Institute IPEV under grant RAYCO, the Swiss National Science Foundation, grants 200020-105435/1 and 200020-113704/1, by the Swiss State Secretariat for Education and Research, grant C05.0034, and by the High Altitude Research Stations Jungfraujoch and Gornergrat. The Russian authors' research is supported by the RBRF grant 09-02-011145-a. We thank the investigators of the following other neutron monitor stations for the data that we used for this analysis: Alma Ata, Apatity, Athens, Baksan, Barentsburg, Calgary, Cape Schmidt, Durham, Fort Smith, Hermanus, Inuvik, Irkutsk, Kingston, Kiel, Larc, Lomnick\'y $\check{\rm S}$t\'it, McMurdo, Magadan, Mawson, Mexico City, Moscow, Mt. Aragats, Mt. Washington, Nain, Nor Amberd, Norilsk, Novosibirsk, Newark, Los Cerrillos, Oulu, Potchefstroom, Rome, Sanae, South Pole, Thule, Tibet, Tsumeb, Tixie Bay and Yakutsk. We acknowledge the supply of radio spectral data via the RSTN web site at NGDC, and S. White (UMd College Park) for the related software. We are particularly grateful to the CORONAS/SONG team (Co PI K. Kudela, Institute of Experimental Physics, Slovakia Academy of Sciences) and D. Haggerty (APL Laurel) for the CORONAS/SONG $\gamma$-ray and ACE/EPAM electron data, and to K.~Shibasaki (Nobeyama Radio Obs.) for the radio polarimetric data and detailed information on their quality. This research has greatly benefitted from the project "Transport of energetic particles in the inner heliosphere" led by W.~Droege at the International Space Science Institute (ISSI) in Bern. Comments on an early version of this manuscrit by G.~Trottet were highly appreciated, as well as discussions with N.~Vilmer, G.~Aulanier and S.~Hoang. S.M.'s doctoral thesis research at Meudon Observatory is funded by a fellowship of Direction G\' en\' erale \`a l'Armement (DGA). 
\end{acks}

 \end{article} 
\end{document}